\documentclass[sigconf,nonacm]{acmart}

\usepackage[utf8]{inputenc}

\usepackage{todonotes}
\usepackage{listings}
\usepackage[normalem]{ulem}

\definecolor{mylightgray}{rgb}{0.97,0.97,0.97}
\definecolor{mygreen}{rgb}{0,0.6,0}
\definecolor{mygray}{rgb}{0.5,0.5,0.5}
\definecolor{mymauve}{rgb}{0.58,0,0.82}

\lstdefinelanguage{hfc}{
	basicstyle=\ttfamily, 
	frame=single,
	basewidth=0.5em,
	sensitive=true,
	morestring=[b]",
	morecomment=[l]{//},
	morecomment=[n]{/*}{*/},
	commentstyle=\color{OliveGreen},
	keywordstyle=\color{blue}\textbf, keywords={def}, otherkeywords={=>,|,\&,!,<=,==},
	keywordstyle=[2]\color{red}\textbf, keywords=[2]{rep,nbr,if,else,share,let},
	keywordstyle=[3]\color{violet}, keywords=[3]{mux,dist,lag,sumHood,countHood,allHood,anyHood,locHood,minHood,maxHood,foldHood,fst,snd,uid,max},
	keywordstyle=[4]\color{orange}\textbf, keywords=[4]{spawn},
	keywordstyle=[5]\color{blue}, keywords=[5]{false,true,infinity,null}
}

\lstdefinelanguage{cpp}{
	basicstyle=\ttfamily\scriptsize,
	keywords={typename,auto,using,namespace,include,class,struct,template,public,return,if,for,while},
	keywordstyle=\color{blue}\textbf,
	otherkeywords={=>,<-,<\%,<:,>:,\#,@},
	keywordstyle=[2]\color{Emerald},
	keywords=[2]{double,bool,int,T,Ts},
	keywordstyle=[3]\color{Brown},
	keywords=[3]{true,false},
	keywordstyle=[4]\color{violet},
	keywords=[4]{MAIN},
	sensitive=true,
	morecomment=[l]{//},
	morecomment=[n]{/*}{*/},
	commentstyle=\color{OliveGreen},
	morestring=[b]",
	morestring=[b]',
	morestring=[b]"""
}

\newcommand{\BNFcce}{{\bf ::=}}
\newcommand{\BNFmid}{\;\bigr\rvert\;}

\newcommand{\e}{\mathtt{e}}

\newcommand{\xname}{\mathtt{x}}
\newcommand{\anyvalue}{\mathtt{v}}

\newcommand{\nbrK}{\mathtt{nbr}}

\newcommand{\neigh}{\rightsquigarrow}

\newcommand{\aEventS}[0]{\mathbb{E}}
\newcommand{\EventS}[0]{\mathbf{E}}
\newcommand{\eventS}[0]{E}
\newcommand{\eventId}[0]{\epsilon}

\newcommand{\deviceId}{\delta}
\newcommand{\sem}[2]{#2 \llbracket {#1} \rrbracket}

\newcommand{\ap}[1]{\langle #1 \rangle}

\newcommand{\formula}[0]{\phi}
\newcommand{\formulalt}[0]{\psi}
\newcommand{\prop}[0]{q}

\DeclareMathOperator{\DX}{\mathrm{Y}}
\DeclareMathOperator{\AX}{\mathrm{AY}}
\DeclareMathOperator{\EX}{\mathrm{EY}}
\DeclareMathOperator{\DU}{\mathrm{S}}
\DeclareMathOperator{\AU}{\mathrm{AS}}
\DeclareMathOperator{\EU}{\mathrm{ES}}
\DeclareMathOperator{\DF}{\mathrm{P}}
\DeclareMathOperator{\AF}{\mathrm{AP}}
\DeclareMathOperator{\EF}{\mathrm{EP}}
\DeclareMathOperator{\DG}{\mathrm{H}}
\DeclareMathOperator{\AG}{\mathrm{AH}}
\DeclareMathOperator{\EG}{\mathrm{EH}}

\usepackage{cleveref}
\usepackage{dblfloatfix}

\AtBeginDocument{%
  \providecommand\BibTeX{{%
    \normalfont B\kern-0.5em{\scshape i\kern-0.25em b}\kern-0.8em\TeX}}}

\setcopyright{acmcopyright}
\acmPrice{15.00}
\acmDOI{XXX}
\acmYear{2022}
\copyrightyear{2022}
\acmSubmissionID{XXX}
\acmISBN{978-1-4503-8546-6/21/07}
\acmConference[VORTEX '22]{Proceedings of the 6th ACM International Workshop on Verification and mOnitoring at Runtime EXecution}{June 6, 2022}{Berlin, Germany}
\acmBooktitle{Proceedings of the 6th ACM International Workshop on Verification and mOnitoring at Runtime EXecution (VORTEX '22), June 6, 2022, Berlin, Germany}

\begin{document}

\title{Predictive Semantics for Past-CTL Runtime Monitors}

\author{Giorgio Audrito}
\orcid{0000-0002-2319-0375}
\affiliation{%
  \institution{Dipartimento di Informatica, \\ Università degli Studi di Torino}
  \streetaddress{Corso Svizzera 185}
  \city{Turin}
  \country{Italy}}
\email{giorgio.audrito@unito.it}

\author{Volker Stolz}
\orcid{0000-0002-1031-6936}
\affiliation{%
  \institution{Høgskulen på Vestlandet}
  \city{Bergen}
  \country{Norway}}
\email{volker.stolz@hvl.no}

\author{Gianluca Torta}
\orcid{0000-0002-4276-7213}
\affiliation{%
  \institution{Dipartimento di Informatica, \\ Università degli Studi di Torino}
  \streetaddress{Corso Svizzera 185}
  \city{Turin}
  \country{Italy}}
\email{gianluca.torta@unito.it}

\begin{abstract}
    The distributed monitoring of swarms of devices cooperating to common global goals is becoming increasingly important, as such systems are employed for critical applications, e.g., in search and rescue missions during emergencies.
    In this paper, we target the distributed run-time verification of global properties of a swarm expressed as logical formulas in a temporal logic. In particular, for the implementation of decentralized monitors, we adopt the Field Calculus (FC) language, and exploit the results of previous works which have shown the possibility of automatically translating temporal logic formulas into FC programs.
    The main limitation of such works lies in the fact that the formulas are expressed in the past-CTL logic, which only features past modalities, and is therefore ineffective in predicting properties about the future evolution of a system. In this paper, we inject some limited prediction capability into the past-CTL logic by providing an extended semantics on a multi-valued logic, then assessing how this affects the automated translation into field calculus monitors.
\end{abstract}


\maketitle

\section{Distributed Runtime Verification}\label{sec:DistributedRuntimeVerification}

Runtime monitoring is a lightweight verification technique dealing with the observation of a system execution with respect to a specification \cite{DBLP:journals/jlp/LeuckerS09}.
Specifications are usually trace- or stream-based, with events mapped to atomic propositions in the underlying logic of the specification language.
Popular specification languages include regular expressions and the Linear Time Logic (LTL).
Distributed runtime monitoring comprises both monitoring of distributed systems, and using distributed systems for monitoring. Distribution is particularly challenging for verification purposes, as it requires to deal with issues such as synchronisation, communication faults, lack of global time, and so on.

In this paper, we address the design of distributed and decentralised runtime monitors \cite{Francalanza2018}, assuming that every agent of the system (e.g., a swarm of drones) executes independently, and occasionally synchronizes or communicates with other agents via a given communication platform. Following Francalanza et al.'s terminology, we model agents as \textit{processes} and consider any two processes as \textit{remote} to each other. Every process produces a \emph{local trace of events}, which is a sequence of sets of observable values derived from the agent's sensors or behaviour. Agents are allowed to appear or disappear from the system, thus different local traces are never aligned in time, i.e., events in the same position of each trace do not necessarily happen at the same time. Monitors check properties of the system by analysing their traces. We follow an \emph{online} evaluation strategy, where the monitors are executed together with the processes themselves, being hosted at the same location and communicating with neighbour monitors. We assume that every agent is executing the same monitor, and this allows us to connect with the traditional setup of runtime verification despite the distributed setting: from the perspective of a single monitor, a single trace is evaluated, although this trace may contain events from remote nodes. Our approach is able to ignore \emph{failures}, which usually make distributed systems harder to manage: a non-responsive node does not disrupt the distributed monitoring process, although influencing its verdict. We do not explicitly address message corruption or faulty sensors, delegating this issue to integrity measures on the communication layer.

Unlike previous works on distributed runtime verification \cite{DBLP:journals/jlp/FrancalanzaGP13}, we propose an automated synthesis of monitors into the Field Calculus (FC) language from high-level specifications in a logic with time modalities. In particular, building on our previous work \cite{adstv:past-ctl}, we extend the semantics presented there to a multi-valued logic, thus allowing a limited, yet useful, level of future prediction.

After introducing Aggregate Computing (the computational model underpinning FC) in section \ref{sec:AggregateComputing}, we briefly review classic two-valued Past-CTL logic and its translation to FC in section \ref{sec:pastctl}. Section \ref{sec:mvpastctl} introduces the extended multi-valued Past-CTL and its translation to FC, while section \ref{sec:examples} presents some examples of the extended logic in action, before we conclude with section \ref{sec:conc}.

\section{Aggregate Computing}\label{sec:AggregateComputing}

The \emph{aggregate computing} paradigm \cite{BPV-COMPUTER2015} has been proposed as a generalisation of previous approaches to programming swarms of devices. One of its most notable features is the ability to express complex, distributed processes through function composition, which supports reusability of collective adaptive behaviour. This aim is achieved through the notion of \emph{computational field}, defined as a global data structure mapping devices of the distributed system to computational values. Fields can be computed from a set of input fields (e.g., from sensors) either at a low-level, through simple programming language constructs, or at a high-level by composition of general-purpose building blocks of reusable collective behaviour, and can be ultimately fed to actuators to implement full-fledged collective adaptive services.

The \emph{Field Calculus} (FC) \cite{avdpb:TOCLhfc} is a minimal language for expressing aggregate computations over swarms of devices, each capable of asynchronously performing simple local computations, and of interacting with a neighbourhood by local exchanges of messages. The FC provides abstraction mechanisms that avoid the need of explicit management of message exchanges, device position, density, and so on. Notably, every device periodically and asynchronously executes the same program following these steps:
\begin{enumerate}
    \item gather contextual information from sensors, local memory, and recently collected messages (i.e., \emph{neighbouring value}s mapping from neighbour devices $\deviceId$ to values $\anyvalue$);
    \item evaluate the program with the information just gathered as input;
    \item store the program result locally, broadcast it to neighbours, and possibly feed it to actuators
\end{enumerate}

Through the repetitive execution by devices of rounds as above, across space and time, a global behaviour emerges; the whole network can thus be viewed as a single \emph{aggregate machine} equipped with a neighbouring relation. The semantics of FC programs can be given through the classical notion of \emph{event structure} \cite{lamport:events}.

\begin{figure}[t]
	\centering
	\includegraphics[scale=1]{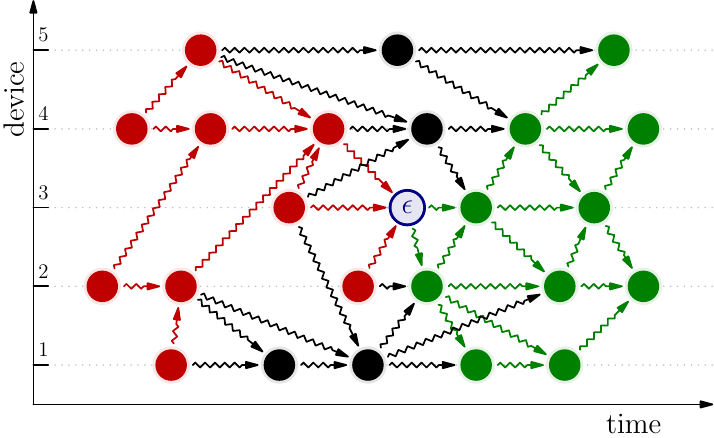}
	\vspace{-10pt}
	\caption{An event structure, split in the causal past of $\eventId$ (red events), causal future (green) and concurrent (black).} \label{fig:structure}
\end{figure}

\begin{definition}[Event Structure] \label{def:eventstruct}
	An \emph{event structure} $\EventS = \ap{\eventS, \neigh}$ is a finite set of events $\eventS$ together with an acyclic \emph{neighbouring relation} $\neigh \subseteq \eventS \times \eventS$ modelling message passing. We say that a sequence of neighbour events $\eventId_1 \neigh \ldots \neigh \eventId_n$ is a \emph{message path}.
\end{definition}
Event neighbouring induces the \emph{causality relation} $<\ \subseteq \eventS \times \eventS$, defined as the transitive closure of $\neigh$ and modelling causal dependence. An example structure is shown in Figure~\ref{fig:structure}. In practice, event structures arise from device neighbourhood graphs changing over time. For instance, device $3$ in Figure~\ref{fig:structure} appears at a certain point in time, with devices 4 and 1 as neighbours, but after a few steps its neighbours become devices 2 and 4. 

In the following, we will present some small snippets of field calculus code, exploiting standard programming language notation together with the following two domain-specific constructs.
\begin{itemize}
    \item
    \lstinline[mathescape]/nbr($\e_0$){$\e$}/.  
    Each device $\deviceId$ evaluates this expression by broadcasting the value of $\e$ to neighbours, and producing a neighbouring value mapping each neighbour $\deviceId'$ of $\deviceId$ (including $\deviceId$ itself) to the latest value that $\deviceId'$ has shared for $\e$. If this is the first execution of the expression, there is no previously shared value for $\deviceId$, and the value of $\e_0$ is used in its place.
    For example, in program \lstinline[mathescape]/nbr(false){$\prop$()}/,
    $\prop$ is a Boolean function returning an observable. In the event $\eventId$ of Figure \ref{fig:structure}, the program computes a map associating devices 2, 3 and 4 to their value of $\prop$ in their last red event.
	\item
	\lstinline[mathescape]/share($\e_0$){($\xname$) => $\e$}/.
	In each device, the result of such an expression is obtained by first:
	\begin{itemize}
		\item
		gathering a neighbouring value $n$ (similarly as $\nbrK$);
		\item
		evaluating $\e$ by substituting $n$ to $\xname$, obtaining the overall value $\anyvalue$ for $\e$; and finally
		\item
		broadcasting $\anyvalue$ to neighbours, which will use it in their next rounds to produce their neighbouring value $n$.
	\end{itemize}
\end{itemize}

For example, consider the following function declaration:
\begin{lstlisting}
def ever(q) {
  share (false) { (n) => anyHood(n) || q }
}
\end{lstlisting}
Function \lstinline/ever/ computes whether a boolean proposition \lstinline/q/ has ever held in the past, through a single \lstinline/share/ construct. This construct gathers neighbours' boolean estimates into a neighbouring value $n$, which is then substituted for variable \lstinline/n/. In the first round of execution, the boolean \lstinline/false/ is used for the current device. The \lstinline/share/ body returns true if \lstinline/q/ holds now, or any neighbour has already recognised that the proposition held in its past, thus realising a gossip routine.

\begin{figure}[!t]
\centering
\centerline{\framebox[\linewidth]{$
	\begin{array}{l@{\,}c@{\,}l@{\hspace{-2mm}}r}
	    \formula & \BNFcce & \bot \BNFmid \top \BNFmid \prop \BNFmid (\neg \formula) \BNFmid (\formula \!\wedge\! \formula) \BNFmid (\formula \!\vee\! \formula) \BNFmid (\formula \!\Rightarrow\! \formula) \BNFmid (\formula \!\Leftrightarrow\! \formula)
	    & {\footnotesize \mbox{logical}}
		\\[3pt]
		& & \BNFmid (\DF \formula) \BNFmid (\AF \formula) \BNFmid (\EF \formula) \BNFmid (\DG \formula) \BNFmid (\AG \formula) \BNFmid (\EG \formula)
		& {\footnotesize \mbox{temporal}}
		\\[3pt]
		& & \BNFmid (\DX \formula) \BNFmid (\AX \formula) \BNFmid (\EX \formula) \BNFmid (\formula \DU \formula) \BNFmid (\formula \AU \formula) \BNFmid (\formula \EU \formula)
		\\[3pt]
	\end{array}
$}
}
\caption{Syntax of past-CTL.} \label{fig:syntax}
\end{figure}

\begin{figure*}[!b]
    \begin{tabular}{|@{\hspace{6.1mm}}c@{\hspace{6.1mm}}|@{\hspace{2mm}}l@{\hspace{16.7mm}}|@{\hspace{6.1mm}}c@{\hspace{6.1mm}}|@{\hspace{2mm}}l@{\hspace{18.75mm}}|@{\hspace{4.85mm}}c@{\hspace{4.85mm}}|@{\hspace{2mm}}l@{\hspace{23mm}}|@{\hspace{2mm}}c@{\hspace{2mm}}|@{\hspace{2mm}}l@{\hspace{24mm}}|}
    	\hline
    	$\top$ & \lstinline|true| &
    	$\prop$ & \lstinline|q()| &
    	$\neg \formula$ & \lstinline[mathescape]/!${\formula}$/ &
    	$\formula_1 \vee \formula_2$ & \lstinline[mathescape]/${\formula_1}$ | ${\formula_2}$/ \\
    	\hline
    \end{tabular}
    \begin{tabular}{|@{\hspace{5.13mm}}c@{\hspace{5.13mm}}|@{\hspace{2mm}}l@{\hspace{7.05mm}}|@{\hspace{3.77mm}}c@{\hspace{3.77mm}}|@{\hspace{2mm}}l@{\hspace{20.3mm}}|}
        \hline
    	${\DX \formula}$ & \lstinline[mathescape]/def Y(f) { locHood(nbr(false){f}) }/ &
    	${\EX \formula}$ & \lstinline[mathescape]/def EY(f) { anyHood(nbr(false){f}) }/ \\
    	\hline
    \end{tabular}
    \begin{tabular}{|@{\hspace{2mm}}c@{\hspace{2mm}}|@{\hspace{2mm}}l@{\hspace{49.3mm}}|}
    	\hline
    	${\formula_1 \!\DU \formula_2}$ & \lstinline[mathescape]/def S(f1,f2)  { share (false) {(old) => f2 | (f1 & locHood(old))} }/ \\
    	${\formula_1 \!\AU \formula_2}$ & \lstinline[mathescape]/def AS(f1,f2) { share (false) {(old) => f2 | (f1 & allHood(old))} }/ \\
    	${\formula_1 \!\EU \formula_2}$ & \lstinline[mathescape]/def ES(f1,f2) { share (false) {(old) => f2 | (f1 & anyHood(old))} }/ \\
    	\hline
    \end{tabular}
    \\~\\[10pt]
    \begin{tabular}{|@{\hspace{6.1mm}}c@{\hspace{6.1mm}}|@{\hspace{2mm}}l@{\hspace{20.75mm}}|@{\hspace{6.1mm}}c@{\hspace{6.1mm}}|@{\hspace{2mm}}l@{\hspace{18.75mm}}|@{\hspace{4.85mm}}c@{\hspace{4.85mm}}|@{\hspace{2mm}}l@{\hspace{23mm}}|@{\hspace{2mm}}c@{\hspace{2mm}}|@{\hspace{2mm}}l@{\hspace{24mm}}|}
    	\hline
    	$\top$ & $\top$ &
    	$\prop$ & \lstinline|q()| &
    	$\neg \formula$ & \lstinline[mathescape]/!$\formula$/ &
    	$\formula_1 \vee \formula_2$ & \lstinline[mathescape]/$\formula_1$ | $\formula_2$/ \\
    	\hline
    \end{tabular}
    \begin{tabular}{|@{\hspace{2mm}}c@{\hspace{2mm}}|@{\hspace{2mm}}l@{\hspace{10.9mm}}|}
    	\hline
    	${\DX \formula}$ & \lstinline[mathescape]/def Y(f)  { if (locHood(nbr($\bot^\cdot$){f}) >= $\top^\cdot$) {$\top^\cdot$ | f & $\top^-$} else {$\bot^-$ | f & $\bot^\cdot$} }/ \\
    	${\EX \formula}$ & \lstinline[mathescape]/def EY(f) { if (anyHood(nbr($\bot^\cdot$){f}) >= $\top^\cdot$) {$\top^\cdot$ | f} else {$\bot^\cdot$} }/ \\
    	${\formula_1 \!\DU \formula_2}$ & \lstinline[mathescape]/def S(f1,f2)  { share ($\bot^\cdot$) {(old) => f2 |  (f1 & if (locHood(old) >= $\top^\cdot$) {$\top^-$} else {$\bot^-$})} }/ \\
    	${\formula_1 \!\AU \formula_2}$ & \lstinline[mathescape]/def AS(f1,f2) { share ($\bot^\cdot$) {(old) => f2 |  (f1 & if (allHood(old) >= $\top^\cdot$) {$\top^\cdot$} else {$\bot$})} }/ \\
    	${\formula_1 \!\EU \formula_2}$ & \lstinline[mathescape]/def ES(f1,f2) { share ($\bot^\cdot$) {(old) => f2 |  (f1 & if (anyHood(old) >= $\top^\cdot$) {$\top$} else {$\bot^\cdot$})} }/ \\
    	\hline
    \end{tabular}
\caption{Translation of a primitive set of past-CTL operators into field calculus. Original translation into Boolean logic (top) and extended translation into multi-valued logic (bottom).} \label{fig:translation}
\end{figure*}

\section{Past-CTL Temporal Logic}
\label{sec:pastctl}
In order to express and monitor properties of distributed systems evolving in time, the past-CTL temporal logic \cite{past:logics,gmr:past-ltl-syntax} has been shown to be naturally translatable into field calculus monitors \cite{adstv:past-ctl}.
Figure~\ref{fig:syntax} presents the syntax of the past-CTL logic. It is based on atomic propositions $\prop$ representing observables, features usual logical operators as well as temporal modalities. These modalities are almost identical to those in traditional CTL, with two main differences:
	(i) temporal operators are interpreted in the past (and thus their names are changed accordingly), along \emph{message paths} that all happened (and are not alternative realities) --- this ensures that formulas have a definite truth value computable at runtime;
	(ii) there are un-quantified versions of the operators along with quantified versions, which refer to the linear past on a single device (and thus behave as past-LTL operators).

Past-CTL formulas can be interpreted in event structures (Fig.~\ref{fig:structure}), giving a truth value for each event. The modalities take inspiration from the words \emph{Yesterday}, \emph{Since}, \emph{Previously}, \emph{Historically}. We choose $\DX, \EX, \DU, \AU$, and $\EU$ as primitive, with the following informal meaning:
\begin{itemize}
	\item $\DX\formula$ means ``$\formula$ held in the previous event on the same device'';
	\item $\EX\formula$ means ``$\formula$ held in some previous event on any device'' {(i.e., in some neighbouring event)};
	\item $\formula\DU\formulalt$ means ``$\formulalt$ held in some past event on the same device, and $\formula$ has held on the same device since then'';
	\item similarly, $\formula\AU\formulalt$ (resp.~$\formula\EU\formulalt$) means ``for all paths (resp.~for some path) of messages reaching the current event, $\formulalt$ held in some event of the path and $\formula$ has held since then''.
\end{itemize}
The other operators can be derived from them by usual means, through $\AX \formula \triangleq \neg \EX \neg \formula$; $\DF \formula \triangleq \top \DU \formula$ (similarly for $\AF$, $\EF$ with $\AU$, $\EU$); $\DG \formula \triangleq \neg \DF \neg \formula$ (similarly for $\AG$, $\EG$ with $\EF$, $\AF$).

A possible translation of past-CTL formulas into field calculus is shown in Figure \ref{fig:translation} (top), by recursion on sub-formulas. We translate atomic propositions $\prop$ into built-in function calls \lstinline|q()| getting their Boolean value from some external environment, and logical operators into their field calculus representation. We assume that:
\begin{itemize}
    \item \lstinline|nbr| and \lstinline|share| are as in Section \ref{sec:AggregateComputing};
	\item \lstinline|allHood|, \lstinline|anyHood|, \lstinline|locHood| are built-in operators collapsing a Boolean neighbouring value $n$ into the conjunction (resp.~{the} disjunction, {the} local value) of its constituent values.
\end{itemize}

{See \cite{adstv:past-ctl} for more details on the recursive mapping of past-CTL to field calculus.}

\section{Multi-valued Past-CTL}
\label{sec:mvpastctl}
We can inject some limited foresight ability into past-CTL by considering a multi-valued logic. The intention of multi-valued logics in runtime verification is to separate an incomplete decision (``not enough information yet'') from a final verdict (``we have reached a stable verdict''). Due to the modal operators in temporal logics, two additional truth-values are often used: for the $\square$-modality (``always''), a value that indicates that so far the property has not been violated, and conversely for the $\diamond$-modality (``eventually'') that a witness has not been observed yet. For distributed runtime verification, we obtain another dimension that we can adequately cover with additional truth-values by distinguishing whether the current verdict holds only for the current device, or for the entire ``cone'' of future events.

We hence obtain the 6-valued logic $\bot < \bot^- < \bot^\cdot < \top^\cdot < \top^- < \top$, to be interpreted {so} that for any event structure $\aEventS'$ extending the current one $\aEventS$:
\begin{itemize}
	\item $\bot$ means ``every future event $\eventId' \geq \eventId$ in $\aEventS'$ will evaluate the formula as $\bot$'' (similar for $\top$);
	\item $\bot^-$ means ``every future event $\eventId' \geq \eventId$ in $\aEventS'$ on the same device as $\eventId$ will evaluate the formula at most as $\bot^-$'' (similar for $\top^-$);
	\item $\bot^\cdot$ means that the formula is false in the current event and no additional information is available (similar for $\top^\cdot$).
\end{itemize}
Notice that a formula can never reach both $\bot$ and $\top$ in different events. In practice, the interpretation of any formula becomes necessarily $\bot$ or $\top$ if in any of its previous events it is $\bot$ or $\top$. Similarly, the interpretation of any formula becomes necessarily $\bot^-$ or $\top^-$ (in both the two-valued and the multi-valued interpretations) if in the previous event on the same device it is $\bot^-$ or $\top^-$. In the following, we take this behaviour for granted and provide the interpretation of a formula only for events which have no preceding events where the same formula is $\bot$ or $\top$, and have no preceding events on the same device where the same formula is $\bot^-$ or $\top^-$.

{Let $\sem{\formula}{\aEventS}(\eventId)$ be the value computed for the multi-valued past-CTL formula $\formula$ at event $\eventId$ in the event structure $\aEventS$.} A locally\nobreakdash-\hspace{0pt}computable multi-valued semantic interpretation of past-CTL formulas can be defined by refining the usual two-valued semantics with the following observations:
\begin{itemize}
	\item $\sem{\formula_1 \AU \formula_2}{\aEventS}(\eventId) = \sem{\formula_2}{\aEventS}(\eventId)$ if $\sem{\formula_2}{\aEventS}(\eventId) \leq \bot^-$ and $\formula_1 \AU \formula_2$ is false in $\eventId$;
	\item $\sem{\formula_1 \EU \formula_2}{\aEventS}(\eventId) = \sem{\formula_1}{\aEventS}(\eventId)$ if $\sem{\formula_1}{\aEventS}(\eventId) \geq \top^-$ and $\formula_1 \EU \formula_2$ is true in $\eventId$;
	\item $\sem{\formula_1 \DU \formula_2}{\aEventS}(\eventId) = \bot^-$ if $\sem{\formula_2}{\aEventS}(\eventId) \le \bot^-$ and $\formula_1 \DU \formula_2$ is false in $\eventId$, and similarly with $\top^-$ if $\formula_1 \DU \formula_2$ is true;
	\item $\sem{\AX \formula}{\aEventS}(\eventId) = \sem{\formula}{\aEventS}(\eventId)$ if $\sem{\formula}{\aEventS}(\eventId) \leq \bot^-$ and $\AX \formula$ is false in $\eventId$;
	\item $\sem{\EX \formula}{\aEventS}(\eventId) = \sem{\formula}{\aEventS}(\eventId)$ if $\sem{\formula}{\aEventS}(\eventId) \geq \top^-$ and $\EX \formula$ is true in $\eventId$;
	\item $\sem{\DX \formula}{\aEventS}(\eventId) = \bot^-$ if $\sem{\formula}{\aEventS}(\eventId) \leq \bot^-$ and $\DX \formula$ is false in $\eventId$, and similarly with $\top^-$ if $\DX \formula$ is true.
\end{itemize}
The resulting translation into field calculus monitors is shown in Fig.~\ref{fig:translation} (bottom). We assume that operators \lstinline|anyHood|, \lstinline|allHood|, \lstinline|locHood| are extended to work on the extended logic values, respectively collapsing the neighbouring value $n$ into the conjunction (resp.~{the} disjunction, {the} local value) of its constituent values. A conjunction (resp.~disjunction) of extended logic values is computed as the minimum (resp.~maximum) of its arguments. Negation also works as usual; and assuming that the extended logic values are implemented as an enumeration type, $\neg v$ can be computed through the subtraction $\top - v$. The translation of the modalities mirrors closely the one in Boolean logic, incorporating extra terms to implement the observations above. 

As a simple example to illustrate the translation, consider formula ${\DX \formula}$, which becomes function \texttt{Y(f)} in field calculus. According to Fig.~\ref{fig:translation} (bottom), the value of \texttt{Y(f)} depends on whether the previous value of \texttt{f} in the current device was true (i.e., $\geq \top^\cdot$) or false; this is checked in the \lstinline|if| condition that, thanks to \lstinline|locHood| applied to \lstinline|nbr| extracts the previous (yesterday) value of \texttt{f} in the current device. If such a value is false, \texttt{Y(f)} will definitely take a false value, but in the multi-valued logic we need to distinguish between $\bot^-$ and $\bot^\cdot$, which is done with expression \lstinline[mathescape]/($\bot^-$ | f & $\bot^\cdot$)/. The disjunction takes the maximum between the two operands; if \texttt{f} has a value greater or equal to $\bot^\cdot$, the right operand of \lstinline/|/ yields $\bot^\cdot$ (since the disjunction takes the minimum of its operands), and the overall value of \texttt{Y(f)} becomes $\bot^\cdot$ because the first operand of \lstinline/|/ is smaller. If, on the other hand, \texttt{f} has a value less or equal than $\bot^-$, the right operand of \lstinline/|/ yields the value of \texttt{f}, and the overall value of \texttt{Y(f)} becomes $\bot^-$ because the first operand of \lstinline/|/ is larger (or equal). Thus, a locally final decision is made (returning $\bot^-$) provided that the previous value is false, and the current value is less or equal than $\bot^-$.
A similar reasoning applies for $\top$ as well; and there is no option to return a $\top$ or $\bot$ value (as in fact according to the semantics there is never enough information to do so). Note that this is exactly in accordance with the multi-valued semantics $\sem{\DX \formula}{\aEventS}(\eventId)$ discussed above.

Notice that this translation is as efficient as the original one, resulting in a program of complexity proportional to the number of connectives used. Further, it is coherent with the previously given translation: collapsing extended logic values to Boolean values directly converts the extended translation into the original translation. Finally, it can be shown by induction that the extended translation reflects the semantics of logic values, so that in particular:
\begin{itemize}
    \item if a formula evaluates to $\top$ (resp.~$\bot$) in any event, it is going to evaluate to $\top$ (resp.~$\bot$) in any following event;
    \item if a formula evaluates to $\top^-$ (resp.~$\bot^-$) in any event, it is going to evaluate to $\top^-$ (resp.~$\bot^-$) in any following event of the same device.
\end{itemize}
Notice that this property is in fact an instance of \emph{impartiality} \cite{DBLP:journals/jlp/LeuckerS09}: once a decision is made, it cannot be retracted.

\section{Examples}
\label{sec:examples}
Consider a networking scenario with two primitive propositions:
\begin{itemize}
    \item $f$: the system is functional in the event;
    \item $b$: a backup has been provided in the event.
\end{itemize}
As a paradigmatic sample application of the multi-valued semantics, we can consider monitoring whether ``\emph{a backup has been made}''. We can model this sentence as the past-CTL formula $\EF b$, which can be expanded into primitive operators as $\top \EU b$. This formula can be either false so far, as no backup has yet been made, or true forever after a backup is completed. The multi-valued semantics captures this fact, by returning either $\bot^\cdot$ (false so far) or $\top$ (true forever).

Additionally, we could also consider monitoring whether ``\emph{the system is always functional}''. We can model this sentence as the past-CTL formula $\AG f$, which can be expanded into primitive operators as $\neg (\top \EU \neg f)$. Similarly as before, the multi-valued semantics succeeds in capturing the fact that either the formula is $\top^\cdot$ (true so far) or $\bot$ (false forever).

Finally, as a more complex example, we could consider monitoring whether ``\emph{every device is aware of a backup, since the time when it knew that the system was always functional}''. This sentence can be modeled in past-CTL as $(\EF b) \DU (\AG f)$, which is a combination of the previous examples. This formula always starts out as \emph{currently true} ($\top^\cdot$), until either a failure happens or a backup is made. If the failure happens first, the formula will become false forever in the device ($\bot^-$). If instead the backup happens first, the formula will become true forever in the device ($\top^-$). Again, the multi-valued semantics successfully captures this fact and allows the user to recognize it without need of particular logical knowledge and reasoning on his side.

\section{Conclusion}
\label{sec:conc}

In this paper, we presented the temporal logic past-CTL together with its automatic translation into aggregate monitors in field calculus. Then, we discussed how the logic semantics could be extended into a 6-valued logic, to inject some limited foresight ability, while retaining the translatability into efficient, resilient and lightweight aggregate monitors. Finally, examples show how this extended semantic can help capturing the future behaviour of a system in some scenarios.
In the future, we plan to test this approach on a simulated realistic case study (such as the ones presented in \cite{ADGMPSTTT:built:environment,DBLP:conf/acsos/AudritoCT21,DBLP:conf/acsos/AudritoCT21a}), and expand on this investigation by adding future modalities to the logic itself, possibly also integrating it with spatial logics as outlined in \cite{at:spacetimelogic}.

\bibliographystyle{ACM-Reference-Format}
\bibliography{long}

\end{document}